\pdfoutput=1
\documentclass[conference]{IEEEtran}
\usepackage[utf8]{inputenc}
\IEEEoverridecommandlockouts
\usepackage{cite}
\usepackage{amsmath,amssymb,amsfonts}
\usepackage{algorithmic}
\usepackage{graphicx}
\usepackage{textcomp}
\usepackage{hyperref}
\usepackage{stfloats}
\usepackage{booktabs}
\def\BibTeX{{\rm B\kern-.05em{\sc i\kern-.025em b}\kern-.08em
    T\kern-.1667em\lower.7ex\hbox{E}\kern-.125emX}}
    
\begin{document}

\title{Cross-paradigm pretraining of convolutional networks improves intracranial EEG decoding\
\thanks{This work was supported by DFG grant EXC1086 BrainLinks-BrainTools and Baden-Württemberg Stiftung grant BMI-Bot.}
}

\author{
    
    \IEEEauthorblockN{Joos Behncke}
    \IEEEauthorblockA{\textit{Department of Computer Science}\\
    \textit{Albert-Ludwigs-University}\\
    Freiburg, Germany\\
    joos.behncke@uniklinik-freiburg.de}
    
    \and
    
    \IEEEauthorblockN{Robin T. Schirrmeister}
    \IEEEauthorblockA{\textit{Translational Neurotechnology Lab}\\
    \textit{University Medical Center Freiburg}\\
    Freiburg, Germany\\
    robin.schirrmeister@uniklinik-freiburg.de}
    
    \and
    
    \IEEEauthorblockN{Martin Völker}
    \IEEEauthorblockA{\textit{Translational Neurotechnology Lab}\\
    \textit{University Medical Center Freiburg}\\
    Freiburg, Germany\\
    martin.voelker@uniklinik-freiburg.de}
    
    \and
    
    \IEEEauthorblockN{Ji\v{r}\'{i} Hammer}
    \IEEEauthorblockA{\textit{Department of Neurology}\\
    \textit{Motol University Hospital, Charles University}\\
    Prague, Czech Republic\\
    jiri.hammer@lfmotol.cuni.cz}
    
    \and
    
    \IEEEauthorblockN{Petr Marusi\v{c}}
    \IEEEauthorblockA{\textit{Department of Neurology}\\
    \textit{Motol University Hospital, Charles University}\\
    Prague, Czech Republic\\
    petr.marusic@lfmotol.cuni.cz}
    
    \and
    
    \IEEEauthorblockN{Andreas Schulze-Bonhage}
    \IEEEauthorblockA{\textit{Epilepsy Center}\\
    \textit{University Medical Center Freiburg}\\
    Freiburg, Germany\\
    andreas.schulze-bonhage@uniklinik-freiburg.de}
    
    \and
    
    \IEEEauthorblockN{Wolfram Burgard}
    \IEEEauthorblockA{\textit{Department of Computer Science}\\
    \textit{Albert-Ludwigs-University}\\
    Freiburg, Germany\\
    burgard@informatik.uni-freiburg.de}
    
    \and

    \IEEEauthorblockN{Tonio Ball}
    \IEEEauthorblockA{\textit{Translational Neurotechnology Lab}\\
    \textit{University Medical Center Freiburg}\\
    Freiburg, Germany\\
    tonio.ball@uniklinik-freiburg.de}
}

\maketitle

\begin{abstract}
When it comes to the classification of brain signals in real-life applications, the training and the prediction data are often described by different distributions. Furthermore, diverse data sets, e.g., recorded from various subjects or tasks, can even exhibit distinct feature spaces. The fact that data that have to be classified are often only available in small amounts reinforces the need for techniques to generalize learned information, as performances of brain-computer interfaces (BCIs) are enhanced by increasing quantity of available data. In this paper, we apply transfer learning to a framework based on deep convolutional neural networks (deep ConvNets) to prove the transferability of learned patterns in error-related brain signals across different tasks. The experiments described in this paper demonstrate the usefulness of transfer learning, especially improving performances when only little data can be used to distinguish between erroneous and correct realization of a task. This effect could be delimited from a transfer of merely general brain signal characteristics, underlining the transfer of error-specific information. Furthermore, we could extract similar patterns in time-frequency analyses in identical channels, leading to selective high signal correlations between the two different paradigms. Classification on the intracranial data yields in median accuracies up to $(81.50 \pm 9.49)\,\%$. Decoding on only $10\%$ of the data without pre-training reaches performances of $(54.76 \pm 3.56)\,\%$, compared to $(64.95 \pm 0.79)\,\%$ with pre-training.
\end{abstract}


\section{Introduction}

After revolutionizing fields like computer vision, deep learning methods have also recently been used to improve classification in applications based on brain computer interfaces (BCIs) \cite{lotte2007review}. A deep belief network model was used to distinguish motor imagery tasks \cite{an2014deep}, outperforming support vector machines (SVM) \cite{chang2011libsvm}, or to extract features of EEG signals \cite{ren2014convolutional}. Other approaches to decode EEG data e.g. used deep convolutional neural networks (deep ConvNets) for feature extraction and visualization \cite{hartmann2018hierarchical}, or built a recurrent convolutional neural network architecture to model cognitive events from EEG data \cite{bashivan2015learning}, applying multi-dimensional features. Likewise for intracranial EEG data, deep neural networks supported classification of epileptic signals \cite{ahmedt2018deep,antoniades2017detection,hosseini2017optimized}.

However, performances of deep learning methods are strongly dependent on the amount of available data. Furthermore, the different methods are mostly restricted to certain conditions when it comes to the design of the data. Assumptions like equal underlying distributions or feature spaces may pertain in classical image recognition tasks, but are mostly not satisfied for real-world applications based on human brain signals. Intra- and inter-individual varieties cause conditions where performances of exactly the same classifier change daily. Also quite similar tasks can exhibit completely different efficiencies in distinguishing classes. In fields such as computer vision, deep learning methods have been enhanced by approaches for transfer learning \cite{pan2010survey,shin2016deep}, especially when only small data are given to train a network. Models, pretrained upon extensive databases \cite{deng2009imagenet}, built the foundation for significant enhancements for example in object categorization or image segmentation \cite{girshick2016region,he2014spatial,everingham2010pascal}. The networks seem to learn the fundamental constitution of the training data to utilize the information for classification in other similar sets. Real-life applications subsist in smooth and fast handling, therefore long training periods are unwanted and collecting substantial real-time data goes beyond the constraints of useful application. 

Recently, transfer learning techniques have found their way into the context of BCI implementations \cite{jayaram2016transfer}. Different approaches are applied e.g. to solve a transfer between different types of error-related potentials \cite{kim2016handling} using a linear support vector machine or to find a way to deal with deviation in latencies \cite{iturrate2014latency} or signal variations \cite{iturrate2013task} in brain controlled interfaces, based on linear discriminant analysis (LDA) \cite{blankertz2011single}. Implementations reverting to deep ConvNets already have generalized non-invasive error-related recordings across subjects, without fine-tuning the network again \cite{volker2018deep}. However, there is still little utilization and transfer learning across different error decoding tasks for intracranial human brain data in combination with deep ConvNets has not yet been investigated. 

In this paper, we determine the impact of transfer learning in intracranial brain recordings across two different error tasks. The paradigms may differ slightly in their way to elicit errors, but basically target the same reaction of perceiving self-caused mistakes in instructed tasks. The classification performances are analyzed separately for both data sets and are compared to those gained by algorithms including transfer learning implantations. It becomes apparent that under conditions with few available data, pre-training and subsequent transfer can improve decoding in error-related classifications tasks. 

\begin{figure*}[hb]
\centerline{\includegraphics[width=\textwidth]{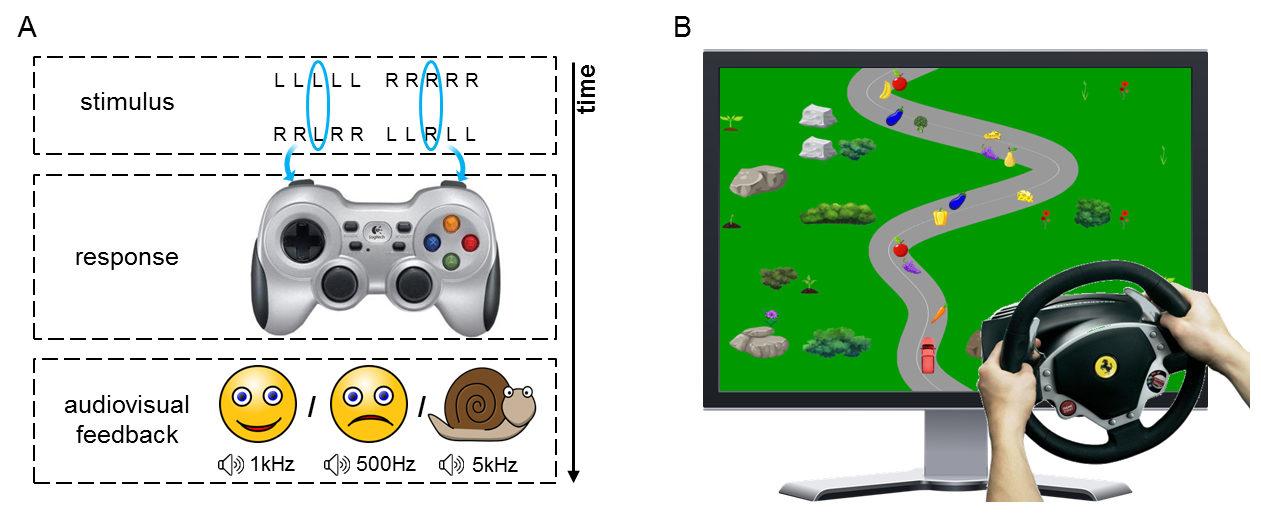}}
\caption{(A) A schematic sketch of the paradigm using an Eriksen flanker task, adapted from \cite{gehring1993neural}. (B) Modified screen shot of the car driving task, in which the participant has to collect rewards and avoid collisions with obstacles (here represented by fruits and vegetables) while keeping the car on the road.}
\label{figtasks}
\end{figure*}

\section{System and Experimental Design}
Two different paradigms were designed to generate a considerable set of intracranial EEG recordings in which error-related brain activity is accessible. In a first paradigm, a flanker task is supposed to elicit error signals under strictly experimental conditions, while in a second paradigm a car game-like environment simulated a more real-life situation. Each participant took part in both experimental paradigms.

\subsection{Eriksen flanker task (EFT)}
This task was designed according to \cite{eriksen1979target}. The participants had to respond to the middle character (either R or L) of a set of letters, acting under time pressure. The audiovisual feedback was given according to a right or a wrong button press, or a reaction slower than a predefined time limit (see figure \ref{figtasks} A). The time limit was set individually to the mean reaction time of a training phase. For details see \cite{volker2018dynamics}. An error was defined as a wrong button press, while a right button press was cited as correct.

\subsection{Car driving task (CDT)}
The second paradigm consisted of a car driving task in which participants were instructed to stay on a road while avoiding certain obstacles (e.g. bombs) punished with a negative score and collecting beneficial objects (e.g. coins) rewarded with plus points (see figure \ref{figtasks} B). As the speed of the game was fixed, the participant’s goal consisted of achieving a highest possible score when reaching the finish line. In this task, an error event was traced when an obstacle was hit; when a beneficial object was hit, the event was declared correct.

\vfill
\section{Pre-processing, Decoding \& Statistics}
The data were obtained by intracranial recordings collected in experiments with 15 patients suffering from epilepsy, who gave their informed consent. According to unique trigger pulses, generated during each experiment, the acquired data were aligned to the event-related meta- information. The aligned data were re-sampled to $250\,Hz$ and re-referenced to a common average, subsequently an electrode-wise exponential moving standardization \cite{schirrmeister2017deep} with decay factor $0.999$ was applied. The data were cut into trials and divided into test and training set according to the specific decoding intervals.

Our decoding architecture made use of the open-source deep learning toolbox \textit{braindecode} for raw time-domain EEG \cite{schirrmeister2017deep}, using the deep ConvNet model \textit{Deep4Net}\footnote{\url{https://robintibor.github.io/braindecode/source/braindecode.models.html}} with a stride of 2. The model comprised four convolution-max-pooling blocks, of these the first block executed step-wise a temporal convolution and a spatial convolution over all channels. Followed by the max pooling, the network owned three conventional convolution-max-pooling blocks. Finally a dense softmax classification layer delivered the output. The network used batch normalization and dropout, exponential linear unit (ELUs) served as activation functions for the different layers. The backward computation of the gradients was based on the output of the categorical cross-entropy loss and optimized using \textit{adam} \cite{kingma2014adam}. Further details of the basic implementation and decisions according to design of the network are discussed in \cite{schirrmeister2017deep}. 

A random permutation test \cite{pitman1937significance} was applied to determine significances per participant. A vector consisting of the true distribution of class labels was compared to $n=10^6$ vectors of randomly shuffled labels to generate a realistic distribution of possible outcomes of the classification. It appeared that the numbers of trials per class were highly unbalanced for all participants. To overcome this problem when creating batches during the training, a class balanced batch size iterator related the samples per batch with the inverse relation to the distribution of the actual trials. For the significance, we solved the imbalance by defining the label matches per vector separately for each class, then averaging over classes and comparing the outcome to the decoded accuracy to estimate the p-value relating to the underlying distribution. Significance was tested for each participant and set of decoding parameters. Single sets exceeding a value of $p=0.05$ were disregarded in further analysis and did not contribute to final results. The significances of the group differences in figure \ref{figtransacc} were determined on the level of trials, using a sign test \cite{hollander2013nonparametric}.

\section{Decodability of Error-related Signals}

\begin{figure}[ht]
\centerline{\includegraphics[width=0.5\textwidth]{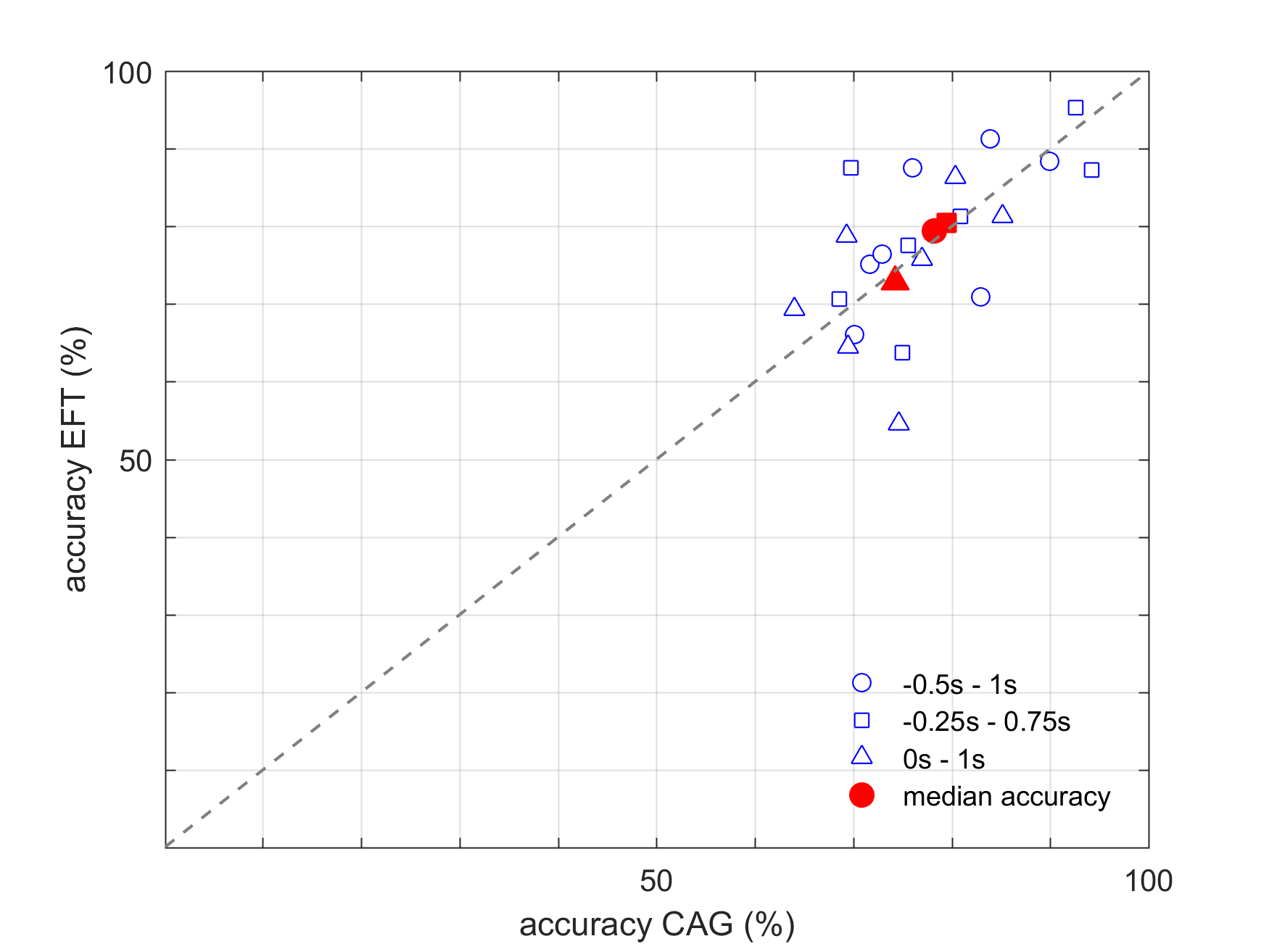}}
\caption{Single participant deep ConvNet accuracies contrasted for the two paradigms and different decoding intervals. Median accuracies per interval are depicted by filled red symbols.}
\label{figallacc}
\end{figure}

For the two data sets, we used the deep ConvNet model \textit{Deep4Net} to determine the two-class decodability of perceived erroneous/correct events in intracranial human brain recordings. Here and in the following, the available data were split into two sets with a proportion of $80\,\%$ for training and $20\,\%$ for testing. For each participant, the decoding accuracy was calculated for different intuitive decoding intervals, which are defined according to the appearance of an event. Figure \ref{figallacc} shows the comparison of the single accuracies for different intervals in blue symbols contrasted for the two data sets and depicts in addition the median accuracy over all participants per interval in form of filled red symbols. In this illustration, only participants are considered who showed significant classification results for both paradigms. 

The classification yielded in median performances of $(78.21  \pm 7.45)\,\%$ for the car driving task and $(79.39  \pm 9.69)\,\%$ for the Eriksen flanker task using the decoding interval $[-0.5, 1]\,s$, $(79.51  \pm 10.33)\,\%$ and $(80.46  \pm 10.81)\,\%$ for the interval $[-0.25, 0.75]\,s$ and finally $(74.27  \pm 7.25)\,\%$ and $(72.91  \pm 10.91)\,\%$ for the interval $[0, 1]\,s$. For both tasks, the interval $[-0.25, 0.75]\,s$ outperformed the others and was therefore used predominantly for the later implementations to transfer learned features. Table \ref{tabcomp} gives an overview of decoding on various intervals and different number of training epochs.

\begin{table}[hb]
\caption{Median Accuracies for Different Decoding Intervals}
\begin{center}
\begin{tabular}{cccc}
\toprule
CDT & & &\\
\midrule
\textbf{epochs}&\textbf{-0.5 - 1s}&\textbf{-0.25 - 0.75s}&\textbf{0 - 1s}\\
\midrule
10 & (72.05 ± 3.28)\,\% & (67.63 ± 3.12)\,\% & (71.27 ± 6.29)\,\% \\ 
50 & (72.91 ± 2.15)\,\% & (73.36 ± 5.00)\,\% & (73.43 ± 1.87)\,\% \\ 
200 & (75.44 ± 3.10)\,\% & (76.94 ± 2.17)\,\% & (74.01 ± 1.99)\,\% \\ 
\midrule
\midrule
EFT & & &\\
\midrule
\textbf{epochs}&\textbf{-0.5 - 1s}&\textbf{-0.25 - 0.75s}&\textbf{0 - 1s}\\
\midrule
10 & (73.71 ± 4.16)\,\% & (82.05 ± 7.85)\,\% & (72.95 ± 6.43)\,\% \\ 
50 & (70.33 ± 5.60)\,\% & (78.47 ± 5.96)\,\% & (70.33 ± 5.60)\,\% \\ 
200 & (81.16 ± 11.10)\,\% & (81.50 ± 9.49)\,\% & (73.56 ± 9.49)\,\% \\ 
\bottomrule
\end{tabular}
\label{tabcomp}
\end{center}
\end{table}

\section{Decoding Transfer Across Paradigms}

\subsection{Responses in the frequency domain}
\begin{figure*}[hb]
\centerline{\includegraphics[width=\textwidth]{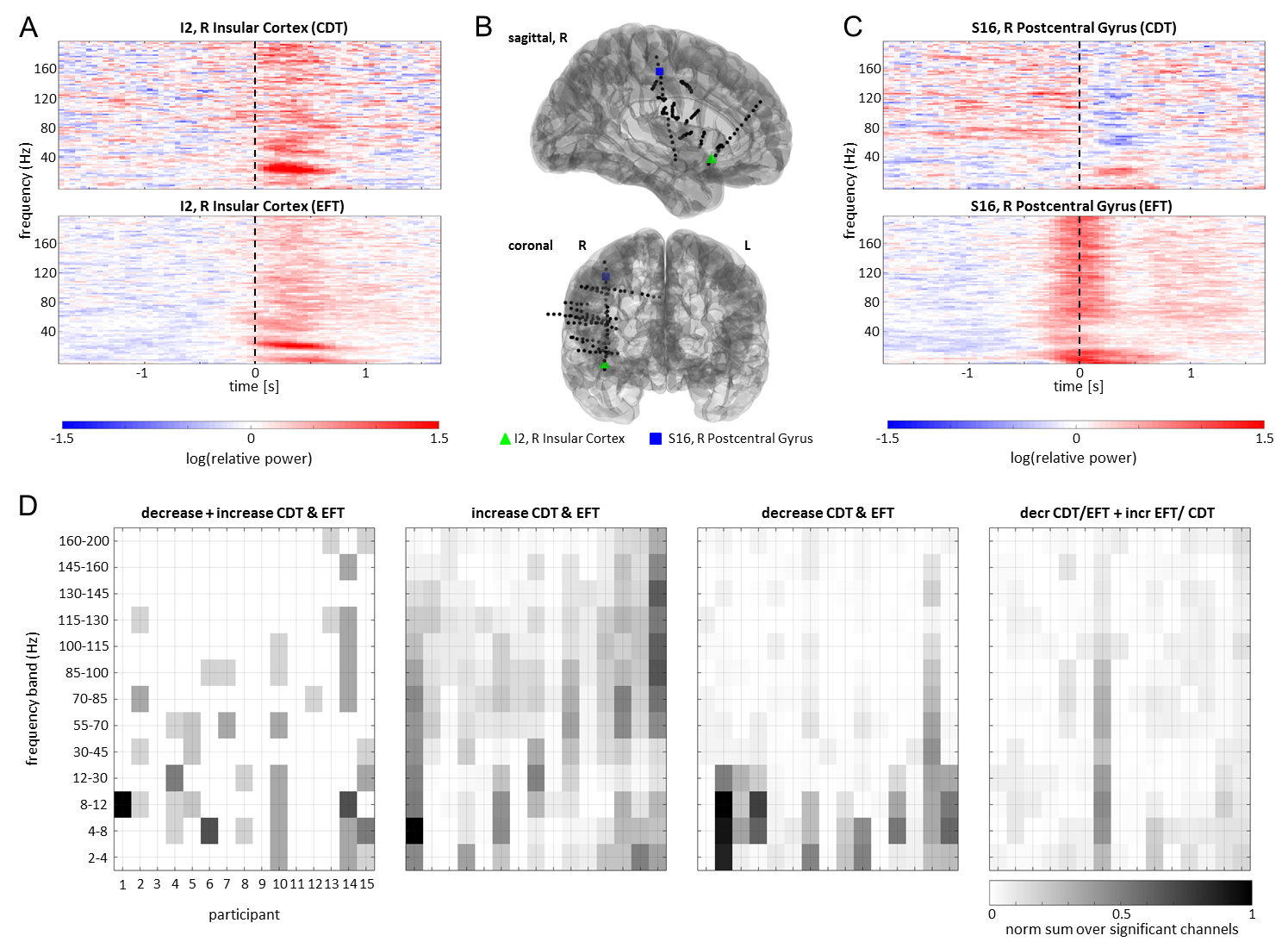}}
\caption{\textbf{Responses in the frequency domain:} \textbf{A} Trial-averaged time-frequency spectra for electrode I2 located in R insular cortex, for error vs correct in CDT (top) and EFT (bottom). \textbf{B} Saggital (top) and coronal (bottom) view of the implanted electrodes for an exemplary participant, plotted on the ICBM152 brain \cite{mazziotta1995probabilistic}. \textbf{C} Trial-averaged time-frequency spectra for electrode S16 located in R postcentral gyrus, for error vs correct in CDT (top) and EFT (bottom). \textbf{D} Normalized sum over significant channels per frequency band and participant, itemized into decrease and increase.}
\label{figallpats}
\end{figure*}

We investigated the single data sets in the frequency domain, using a multitaper method to estimate the power spectral density \cite{thomson1982spectrum}. Optical inspection and comparison of time-frequency spectra for identical electrodes but different tasks revealed obvious similarities for several electrodes. Figure \ref{figallpats}A depicts one example where a resemblance is unambiguous, showing the response for error vs correct in electrode I2 located in the right insular cortex. The dotted line marks the onset of the event. Nevertheless, other electrodes did not show any effects, or effects could only be seen strongly for one of the tasks, as illustrated in figure \ref{figallpats}C. A global overview of all electrodes for this exemplary participant is given in figure \ref{figallpats}B. The blue and green markers refer to the electrodes selected for figures \ref{figallpats}A and \ref{figallpats}C.

Moreover, we analyzed the behaviour of frequency-band power time-series of significant channels. Decrease and increase of the power were tagged for both paradigms, CDT and EFT, and compared among themselves, to get an estimation of similarities in temporal developments of the frequency power. Figure \ref{figallpats}D illustrates the outcome of this type of analysis, dividing the figure into four conditions of overlapping tags for the two paradigms. The color code in each subplot refers to the sum of significant channels, normalized to the number of channels per participant and to the maximal value of significant channels, exhibiting the specific tag indicated by the subplot title. The individual color values are broken down to frequency band and participant. Significant decrease for both paradigms as well as a significant increase in the lower frequency bands ($< 30\,Hz$) can be seen in the data for most of the participants. However, for all participants an increase in the gamma band is prominent, covering the bands from $55\,Hz$ to $130\,Hz$. For some of the participants the manifestation is present in more channels than for others, according to the specific implantation and the adjacent brain area.

\subsection{Compilation of different transfer techniques}
Initially we examined the output of three different transfer techniques, choosing a small number of post-training epochs compared to the number of epochs ($n=200$) in the pre-training with the first data set. An assembly of the results is given in table \ref{tabcomptrans}, showing the median accuracy over the participants. Errors were estimated by selecting the interquartile range of the bootstrapped samples per interval and technique. For each of the three implementations, the network was pre-trained on a given data set $D_i$, while a then unknown set $D_j$ was used for testing or fine-tuning, respectively. The whole data were processed so that the feature space remained the same for the two sets. Therefore an adjustment of the input layer was not necessary. 

\begin{table*}[ht]
\caption{Median Accuracies for Different Transfer Techniques}
\begin{center}
\begin{tabular}{ccccc}
\toprule
\multicolumn{3}{r}\textbf{fine-tuning on $D_{CDT}$ (network pre-trained on $D_{EFT}$)} & \\
\midrule
\textbf{layers}&\textbf{epochs}&\textbf{-0.5 - 1s}&\textbf{-0.25 - 0.75s}&\textbf{0 - 1s}\\
\midrule
all & 0 & (50.46 ± 1.11)\,\% & (49.28 ± 0.65)\,\% & (48.82 ± 2.04)\,\% \\
all & 10 & (67.53 ± 1.39)\,\% & (66.84 ± 10.23)\,\% & (69.78 ± 3.32)\,\% \\ 
last & 50 & (61.32 ± 2.87)\,\% & (62.98 ± 1.50)\,\% & (63.01 ± 5.82)\,\% \\ 
\midrule
\midrule
\multicolumn{3}{r}\textbf{fine-tuning on $D_{EFT}$ (network pre-trained on $D_{CDT}$)} & \\
\midrule
\textbf{layers}&\textbf{epochs}&\textbf{-0.5 - 1s}&\textbf{-0.25 - 0.75s}&\textbf{0 - 1s}\\
\midrule
all & 0 & (53.98 ± 4.90)\,\% & (57.46 ± 6.70)\,\% & (54.19 ± 3.56)\,\% \\ 
all & 10 & (73.44 ± 7.94)\,\% & (72.12 ± 7.92)\,\% & (76.83 ± 13.47)\,\% \\ 
last & 50 & (66.67 ± 4.02)\,\% & (68.89 ± 2.86)\,\% & (59.48 ± 6.64)\,\% \\
\bottomrule
\end{tabular}
\label{tabcomptrans}
\end{center}
\end{table*}

The first approach consisted of the pre-training and subsequently classification on the second unseen set $D_j$ based on the pre-defined weights without fine-tuning. Generalizing from EFT to CDT, the deep ConvNet was not able to predict the true classes of the tasks and presented poor performances around chance. For the transfer from the CDT to the EFT data set accuracies were slightly better, exceeding chance level and showing a peak performance of $(57.46  \pm 6.70)\,\%$ for the interval $[-0.25, 0.75]\,s$.

Secondly, the pre-trained network was fine-tuned by training on a then unknown data set $D_j$ for $n=10$ epochs with a smaller learning rate. Here indeed the network learns informative features and obtains accuracies around $70\,\%$ for both of the paradigms. However, comparison with the performances given in table \ref{tabcomp} indicates that there is no enhancement when using the pre-training. To the contrary, the accuracies do not yield the high values obtained by training directly on the classification data set training for $n=10$ epochs.

The third implementation was inspired by techniques from computer vision, where networks are pre-trained by a huge training set and only a few last layers are trained again by a smaller set of similar data to fine-tune the weights in the deeper layers. We captured this idea and froze all layers after pre-training and adjusted only the weights of the last classification layer. In both data sets performances yielded accuracies of $60\,\%$ and higher, but not reaching the values obtained when fine-tuning the whole network, even with less epochs.

\subsection{Performance dependency on the amount of data}
\begin{figure*}[ht]
\centerline{\includegraphics[width=0.8\textwidth]{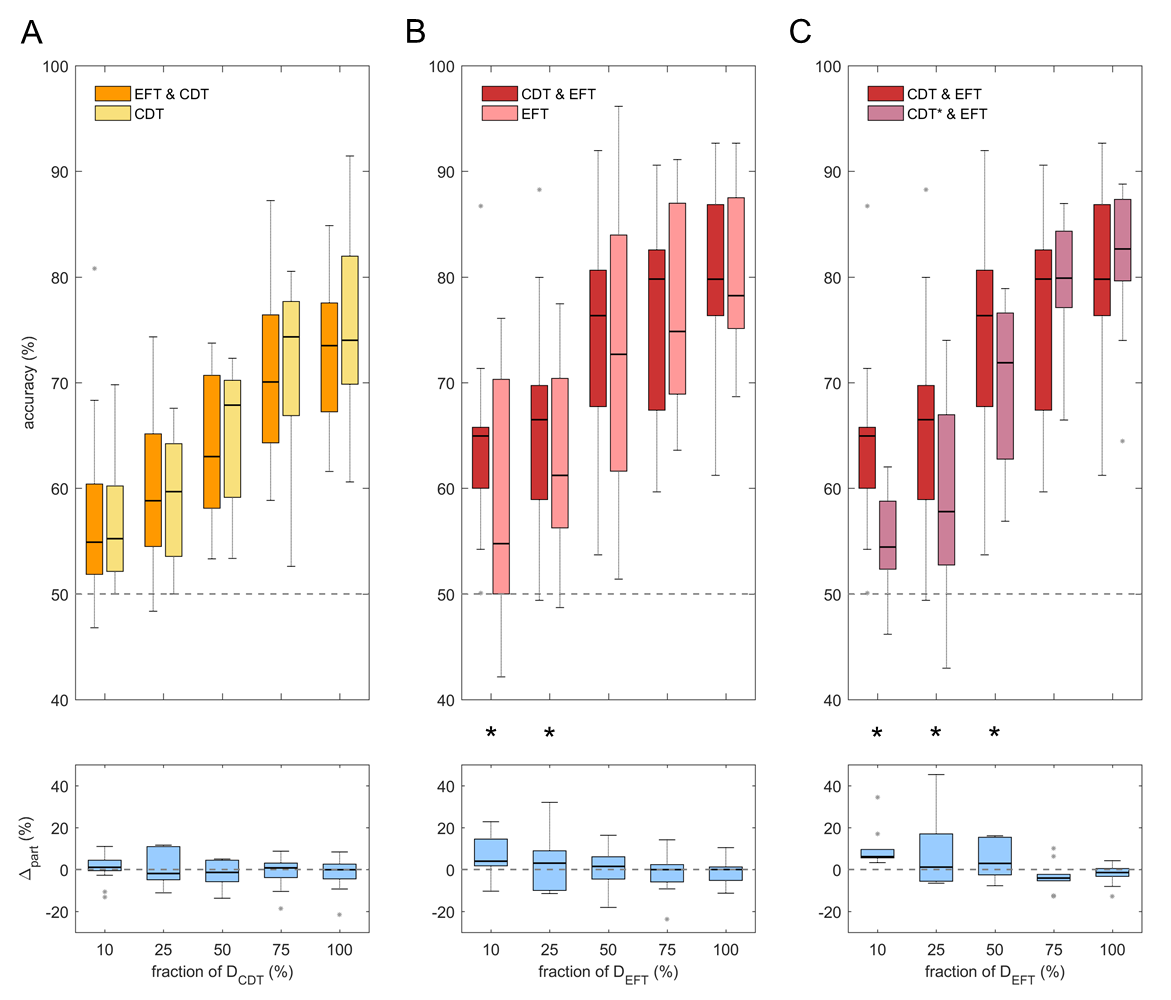}}
\caption{\textbf{Contrast of median accuracies for vanishing data:} Accuracies obtained by stepwise reduction of available training data $D_{CDT}$, comparing (a) the training only on $D_{CDT}$ to (b) pre-training on $D_{EFT}$ and then fine-tuning on $D_{CDT}$ (\textbf{A}, top) and vice versa for $D_{EFT}$ (\textbf{B}, top). \textbf{C} (top) Pre-training on $D_{CDT}$ and fine-tuning on $D_{EFT}$ compared to pre-training on $D_{CDT*}$ with shuffled labels and fine-tuning on $D_{EFT}$. \textbf{A}-\textbf{C} (bottom) Accuracy distribution of intra participant difference, e.g. $ACC_{transf} - ACC_{CDT}$ for \textbf{A}. * indicates significance of the difference with $p < 0.05$.}
\label{figtransacc}
\end{figure*}

Again, the network was pre-trained on a given data set $D_i$ to implement the weights. To draw a comparison between conditions with only few data and situations where more data are available, we took the second data set $D_j$ and gradually reduced the amount of data used for fine-tuning from $100\,\%$ to $10\,\%$ of the available training data ($80\,\%$ of the entire data), once more employing a smaller learning rate as in the pre-training. Median accuracies and the underlying distributions are presented in figure \ref{figtransacc}A (top) for $D_j = D_{CDT}$ and figure \ref{figtransacc}B (top) for $D_j = D_{EFT}$. The boxes depict the interquartile range, the whiskers extend to the most extreme data points and outliers are drawn as asterisks. The plots at the bottom of each subfigure reveal the distribution of the intra-participant difference between the two compared decoding accuracies. E.g. to obtain the values for figure \ref{figtransacc}A we calculated the difference $ACC_{transf} - ACC$ for each participant, where $ACC_{transf}$ corresponds to the accuracy gained with pre-training while $ACC$ corresponds to the accuracy achieved without pre-training. For decoding on $D_{CDT}$, figure \ref{figtransacc}A, there is no big difference between the two conditions. Even with less data for the final training, the pre-training cannot enhance the performance. In contrast, in figure \ref{figtransacc}B the pre-training on $D_{CDT}$ has the effect that for a decreasing amount of data the performance gradually gets better, exhibiting significant differences of median accuracies up to $10\,\%$ for a fraction of $10\,\%$ of the training data. The distribution of the intra-participant differences for the smaller amount of used data confirm this trend. Median accuracies yielded with pre-training are consistently better than in cases when only the training on the unseen set was performed. Due to the relatively small number of participants, significance was tested on the level of single trials.

A last comparison claims to test whether the distinction between the two cases originates from a transfer of more general features of the brain signals and not the true underlying conditions. Therefore,  the performed transfer was contrasted to the decoding results of pre-training on $D_{CDT}$ with randomly shuffled labels and then fine-tuned on $D_{EFT}$. Hereby the network wasn't able to learn the features of the two conditions. Indeed the results show that the decoding using unshuffled labels during the pre-training performs clearly better for decreasing data, as illustrated in \ref{figtransacc}C. The lower plot again shows the distribution of the intra-participant difference, where the values were determined by $ACC_{transf} - ACC_{shuffle}$. Here, too, differences for the fewer data exhibit positive median values and distributions mainly over zero.

\section{Conclusion \& Outlook}

In this paper, two different issues were analyzed. First, the proof of decodability of error-related signals in the underlying intracranial brain recordings was brought to the fore. This was tested for two paradigms, differing by their affinity to real-life application. Error decoding has been investigated several times using EEG data e.g. when observing and controlling robots \cite{iturrate2015teaching,salazar2017correcting,behncke2018signature}, or in real interaction simulations \cite{buttfield2006towards}, but not yet on the basis of intracranial recordings. Here, we obtained accuracies up to $(79.51  \pm 10.33)\,\%$ for the CDT and $(80.46  \pm 10.81)\,\%$ for the EFT. The quite high performances reinforce the use of these data for approaches reverting to transfer techniques. However, the high errors show non-negligible differences of the results, which certainly should be treated with caution. Different patients were equipped with differing implantations, which in turn covered different brain areas. Thus, it cannot be excluded that more or less informative channels were given in the varying data sets, leading necessarily to diverse decoding performances. Because of the different implantations, we abstained from an inter-subject transfer.

The second aspect concerned the similarity of the data sets gained by the different paradigms and their transferability. Time-frequency spectra of same channels revealed striking similarities for some of the channels. More precise examinations of frequency-band dependent time-series of the power spectral density uncovered an extensive increase of significant channels in the gamma band between $55\,Hz$ and $130\,Hz$, as already indicated in \cite{volker2018dynamics}. Likewise, the results indicate a similarity in the characteristics of the data for the two distinct paradigms.

A comparison of several transfer approaches for the whole extent of data but a lower number of epochs did not lead to improvement of the decoding. When the network was trained directly with the objective data set exclusively, higher accuracies were yielded compared to pre-training the network. As already shown by \cite{volker2018deep} on EEG data, a direct transfer without further fine-tuning did not succeed.

In many cases, acquiring intracranial data is hardly possible and raised data sets are often not extensive. In this paper we illustrated a significant improvement of decoding for decreasing amounts of data when the network is pre-trained by a similar set. Interchanging the two data sets led to no enhancements, which might be explained by the fact that in this case the pre-training was performed on the set comprising only few trials and therefor possibly made the generalities of the conditions not sufficiently or hardly learnable. Instead the question arises whether, for a transfer, the relation of the amount of data used for pre- and post-training plays a determining role for the applicability of this technique. Certainly, a degree of similarity between the data sets has to be given, also with respect to the manifestation of the two conditions, which could be shown here by randomization of labels.

Several interesting questions and approaches can be deduced from these results. E.g. a network might be trained on an extensive set of non-invasive data to learn problem-specific characteristics, which subsequently can be fine-tuned by a small intracranial data set. Here, a change of network architecture can make a transfer possible, assuming data in different feature spaces. Likewise, data augmentation can contribute to advance classification in rather small data sets.

\section*{Acknowledgment}

The authors would like to thank everyone involved in creating and processing the data sets, but especially the patients for their conscientious participation. 

\bibliographystyle{IEEEtran}
\bibliography{IEEEabrv,ms}

\end{document}